\documentclass[review]{elsarticle}

\usepackage{lineno,hyperref}
\usepackage{float}
\modulolinenumbers[5]
\usepackage{amsmath}

\usepackage{graphicx} 
\usepackage{subfigure} 

\begin{document}

\begin{frontmatter}

\title{Photorefractive and computational holography in the experimental generation of Airy beams}





\author[mymainaddress]{Rafael A. B. Suarez}

\author[mymainaddress]{Tarcio A. Vieira}

\author[mymainaddress]{Indira S. V. Yepes}

\author[mymainaddress]{Marcos R. R. Gesualdi\corref{mycorrespondingauthor}}
\cortext[mycorrespondingauthor]{Corresponding author}
\ead{marcos.gesualdi@ufabc.edu.br}

\address[mymainaddress]{Universidade Federal do ABC, Av. dos Estados 5001, CEP 09210-580, Santo Andr\'e, SP, Brazil}

\begin{abstract}
In this paper, we present the experimental generation of Airy beams via computational and photorefractive holography. Experimental generation of Airy beams using conventional optical components presents several difficulties and are practically infeasible. Thus, the optical generation of Airy beams has been made from the optical reconstruction of a computer generated hologram implemented in a spatial light modulators. In the photorefractive holography technique, being used for the first time to our knowledge, the hologram of an Airy beam is constructed (recorded) and reconstructed (reading) optically in a nonlinear photorefractive medium. The Airy beam experimental realization was made by a setup of computational and photorefractive holography using a photorefractive $Bi_{12}TiO_{20}$ crystal as holographic recording medium. Airy beams and Airy beam arrays were obtained experimentally as in accordance with the predicted theory; and present excellent prospects for applications in optical trapping and optical communications systems.
\end{abstract}

\begin{keyword}
\texttt{Airy beams, computational holography, spatial light modulator, photorefractive holography}
\end{keyword}

\end{frontmatter}

\section{Introduction}
In recent years, the study of Airy beams has attracted great interest in optics and atomic physics due to their unusual features such as the ability to remain diffraction-free over long distances while they tend to freely accelerate during propagation,  this was showed theoretically and experimentally by Siviloglou $et$ $al$ \cite{Sivilo2007,Siviloglou2007}. The origin of these strange features, explained by Berry and Balazs in 1979, is due to a non-trivial solution of the Schrodinger equation in quantum mechanics for a free particle and the caustic envelope overlapped by the superposition of a plane wave \cite{Berry1979}. These self-accelerating Airy beams have also inspired prominent research interests and potential applications such as: optical micromanipulation \cite{Zhang2011,Cheng2014,Zheng2011,Cao2011}; plasma physics \cite{Polynkin2009,Klein2012}; optical microscopy \cite{Vettenburg2014}; and, recently, the growing interest in the influence of optical vortices on Airy beams \cite{Dai2010,Jiang2012,Li2014}.  

On the other hand, the holography, which was proposed by Dennis Gabor in 1948 \cite{Gabor1948}, enables the information of the amplitude and the phase of an object or optical wave can be recorded on a holographic recording medium. Since the development of the laser in the 1960s, there have been several works of conventional holography that used a laser as a source of coherent light and holographic recording materials such as silver halide films, thermoplastics, and photosensitive materials. In addition to the conventional holographic method, classical (or analogical) holography, several others were proposed as computational (or numerical) holographic methods: computer generated holograms, when the recording or construction of the hologram is numerical (CGH) \cite{Tricoles1987,Vasara1989}; and, digital holography, when numerical reconstruction of holograms are recorded with CCD sensors of high resolution \cite{John2005,Brito2013,Sutkowski2000}.

Additionally, the development of computers and electronic devices that are ever faster and are of higher resolution, such as CCD cameras and the spatial light modulators (SLMs) based on liquid crystal display (LCD) or micro mirrors (DMDs); new laser sources; optical systems and opto-mechanical devices which are also of excellent quality. This has enabled the experimental implementation of holographic systems of numerical and optical reconstruction of wavefronts of objects and optical beams possible.

Photorefractive holography has been presented as a promising technique for dynamic processing of record-holographic reconstruction and holographic interferometry techniques to analyze surfaces and optical wavefronts \cite{Gunter2006,Gesualdi2006,Gesualdi2007,Gesualdi2008,Gesualdi2010,Brito2013}. This is based on the photorefractive effect, consisting in modulation of the refractive index via photoinduction of charge carriers and linear electro-optic effect in some semiconductor crystals with a particular features, the so-called photorefractive crystals ($LiNbO_{3}$, SBN, KBT, $BaTiO_{3}$, $Bi_{12}TiO_{20}$, among others)\cite{Gunter2006}. Due to the fact that it is a process that occurs at the electronic level of semiconductor crystals with nonlinear optical properties, the holographic networks feature high resolution and short response time, making it possible to act as a holographic recording media that does not require chemical or computational processing for reconstruction of the holographic image and presents indefinite reusability \cite{Gunter2006,Gesualdi2006,Gesualdi2007,Gesualdi2008,Gesualdi2010}.

Based on all this, we present the experimental generation of Airy beams via computational and photorefractive holography. The Airy beam experimental realization was made by a setup of computational and photorefractive holography using a photorefractive $Bi_{12}TiO_{20}$ crystal as holographic recording medium. The Airy beams and  Airy beams arrays obtained experimentally are in accordance with the predicted theoretically; and present excellent prospects for applications in optical trapping  \cite{Zhang2011,Cheng2014,Zheng2011,Cao2011} and optical communications systems \cite{Polynkin2009,Klein2012}.

%
\section{ Airy Beams: theoretical background}
The solution for Airy beams propagating with infinite energy can be obtained by solving the normalized paraxial equation of diffraction in $(1+1)$D \cite{Sivilo2007} 
\begin{equation}
i\frac{\partial}{\partial \xi}\phi\left(s,\xi \right)+\frac{1}{2}\frac{\partial^{2}}{\partial s^{2}}\phi\left(s,\xi \right)=0
\label{paraxial}
\end{equation}
where $\phi$ is the complex amplitude of the electric field associated with planar Airy beams; $s=x/x_{0}$ and  $\xi=z/kx_{0}^{2}$ are the dimensionless transverse and longitudinal coordinates; $x_{0}$ is an arbitrary transverse scale and $k=2\pi n/\lambda_{0}$ is the wavenumber of an optical wave. 

An ideal Airy beam is immune to diffraction effects, but it possesses infinite energy making its experimental generation impossible. However, it is possible to obtain a finite-energy version of the ideal Airy beam by modulating it with a spatial exponential function on the initial plane $z=0$, that is, the scalar field in $\xi=0$ is given by \cite{Sivilo2007,Siviloglou2007,Berry1979}:
\begin{equation}
\phi\left(s,0 \right)=Ai\left(s \right)exp\left(as \right)
\label{Airy_finity}
\end{equation} 
where $Ai$ is the Airy function and $a$ is a positive quantity which ensures the convergence of Eq.\ref{Airy_finity}, thus limiting the infinity energy of the Airy beams. The scalar field $\phi\left(s,\xi \right)$ is obtained from the Huygens-Fresnel integral, which is highly equivalent to Eq.\ref{paraxial} and determines the field at a distance of $z$ as a function of the field at $z=0$, that is
\begin{equation}
\phi\left(s,\xi \right) =Ai\left(s-\frac{\xi^{2}}{4}+ia\xi\right)exp\left(as-\frac{a\xi^{2}}{2}-i\frac{\xi^{3}}{12}+i\frac{a^{2}\xi}{2}+i\frac{s \xi}{2}\right)   
\label{Airy_finity1}
\end{equation}
This equation shows that the intensity profile decays exponentially as a result of modulating it with a spatial exponential function on the initial plane $z=0$. The term $s_{0}=s-\left( \xi^{2}/4\right)$, where $s_{0}$ denotes the initial position of the peak at $z=0$, defines the transverse acceleration of the peak intensity of Airy beams and describes the parabolic trajectory.

These results are generalized in $(2+1)$D taking the scalar field and what described the beam as the product of two independent components \cite{Sivilo2007}, that is:
\begin{equation}
\phi\left(x,y,z\right)=\phi_{x}\left(x,z\right)\phi_{y}\left(y,z\right)
\label{Airy_2D}
\end{equation}
where each of the components satisfies the Eq.\ref{paraxial}. Thus, the initial field to $z=0$ is:
\begin{equation}
\phi\left(x,y,z=0\right)=Ai\left(\frac{x}{x_{0}} \right) Ai\left(\frac{y}{y_{0}} \right)exp\left[\left(\frac{x}{w_{1}} \right)+\left(\frac{y}{w_{2}} \right)  \right] 
\label{Airy_2D_finita}
\end{equation}
%
\section{Generation of Airy beams through computational holographic technique using spatial light modulator (SLM)}
With the development of spatial light modulators (SLM), the generation of special optical beams (non-diffractive or diffraction-resistant) becames possible via holographic techniques due to the ability to individually control each pixel in real time and its high fidelity of signal reconstruction in holographic systems\cite{Vasara1989,Arrizon2005,Vieira2014,Vieira2012}.

Using the fields that described the Airy beams, Eq. \ref{Airy_finity}, we built a computer generated hologram (CGH) that is optically reconstructed using a nematic liquid crystal spatial light modulator(LC-SLM). Due to their simplicity and the fact of not requiring high diffraction efficiency, we used an amplitude hologram implemented on a reflective spatial light modulator. The hologram is implemented using an amplitude function that consists in varying the coefficient of transmission or reflection of the medium as from the following amplitude function \cite{Arrizon2005,Vieira2014,Vieira2012}
\begin{equation}
H\left( x,y\right)= \frac{1}{2}\left\lbrace \beta\left( x,y\right) +a\left( x,y\right)cos\left[\phi\left( x,y\right)-2\pi \left(\xi x + \eta y \right)  \right] \right\rbrace 
\label{Transmissão}
\end{equation}
where $a\left( x,y\right)$ and $\phi\left( x,y\right)$ are the amplitude and phase of the field complex respectively; $\left(\xi,\eta \right)$ is a spacial frequency of the plane wave using as reference and $\beta\left( x,y\right)=\left[1+a^{2}\left(x,y \right)\right]/2 $ is the function bias taken as a soft envelope of the amplitude $a\left( x,y\right)$. The implementation of a pixelated version of the transmittance Eq.\ref{Transmissão} on a SLM presents remarkable advantages as to the precision, as well as the fact that the object beam and the reference beam are mathematical entities that are specified with versatility and precision. The plane wave of reference is off-axis and introduces frequencies that separate the different orders of the encoded field. 
\subsection{Experimental setup}
\begin{figure}[H]
 \centering
  \includegraphics[scale=0.4]{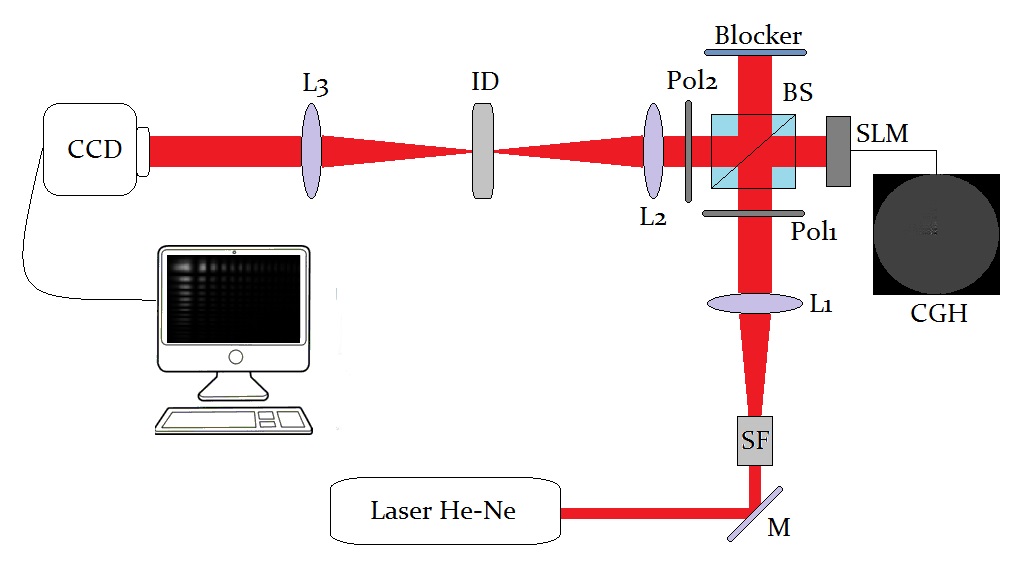}
 \caption{Experimental holographic setup for Airy beams generation, where Ms are mirrors, BS is a beam splitter, FS is the spatial filter, Ls are lenses, Pol’s are polarizers, SLM is spatial light modulator, ID is the mask and CCD camera for image acquisition.}
 \label{arranjo}
\end{figure}

The Fig. \ref{arranjo} shows the experimental holographic setup for the optical reconstruction process of the holograms (CGHs) of an Airy beam \cite{Vieira2012}. The He-Ne laser beam $\left( \lambda=632.8nm\right) $, that is expanded by the spatial filters $SF$ and collimated by the lens $L_{1}$, incident perpendicularly on the surface of the SLM ($LC-R1080$ model, of the Holoeye Photonics, with a liquid crystal display of  $1980\times 1200$ and pixel $8.1\mu m$)  that is placed at the input plane (focus of lens $L_{2}$). In order to obtain an amplitude modulation the optical axis of polarizer $pol_{1}$ is aligned forming an angle of $0^{0}$ and of the polarizer $pol_{2}$ in $90^{0}$ with relation to the axis $y$ of the SLM.  A 4$f$ spatial filtering system, where a spatial filtering mask (ID, bandpass rectangular aperture in the case of $(1+1)$D Airy beam and circular aperture in the case of $(2+1)$D Airy beam) is placed at the Fourier plane, allows selecting and transmission of the light from the signal spectrum generating the encoded signal at the output plane of the holographic setup, the field $\phi\left(x,y,z\right)$. Finally, the transversal intensity pattern of the reconstructed Airy beam is recorded by a CCD camera "step by step" along the propagation axis defined by the distance of propagation invariant finite of the Airy beam.
\subsection{Experimental results}
\begin{figure}[htbp]
\centering
 \subfigure[]{\includegraphics[width=50mm]{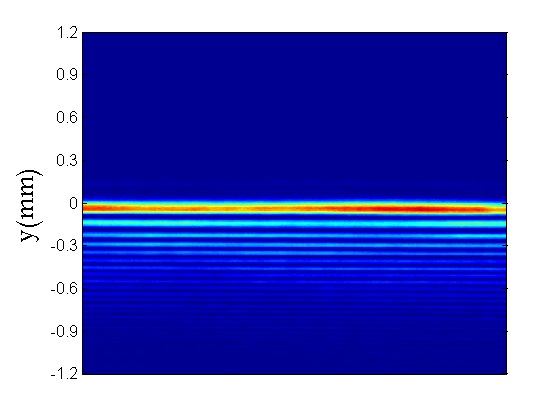}}
 \subfigure[]{\includegraphics[width=50mm]{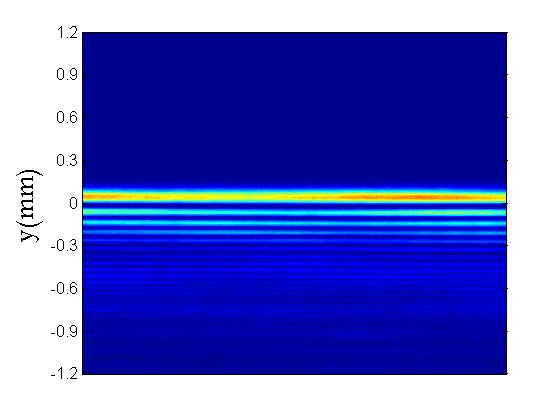}}
 \subfigure[]{\includegraphics[width=50mm]{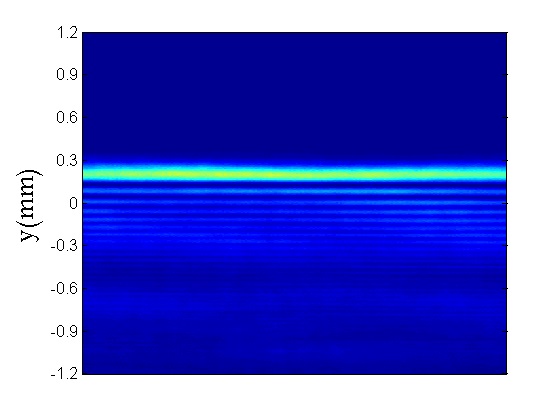}}
 \subfigure[]{\includegraphics[width=50mm]{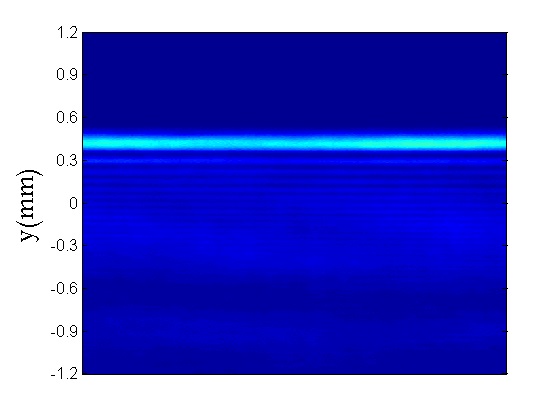}}
\caption{Transversal intensity pattern for an Airy beam in $(1+1)D$ when $x_{0}=50 \mu m$ and $a=0.07$ in $(a)$ $z=3cm$, $(b)$ $z=7cm$, $(c)$ $z=11cm$, $(d)$ $z=15cm$.}
 \label{Transversal_1D_50_0.07_1}
\end{figure}

\begin{figure}[H]
\centering
 \subfigure[]{\includegraphics[width=50mm]{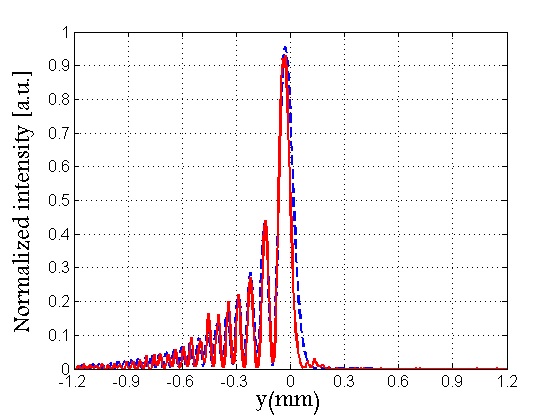}}
 \subfigure[]{\includegraphics[width=50mm]{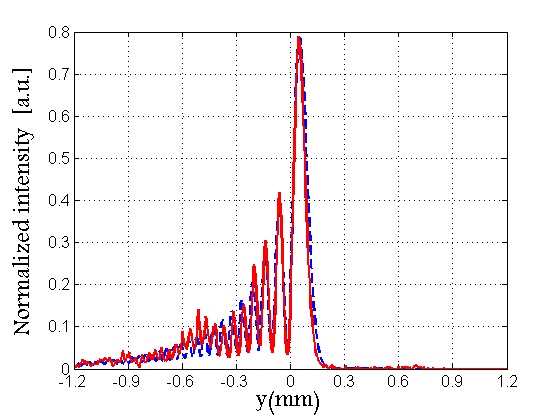}}
 \subfigure[]{\includegraphics[width=50mm]{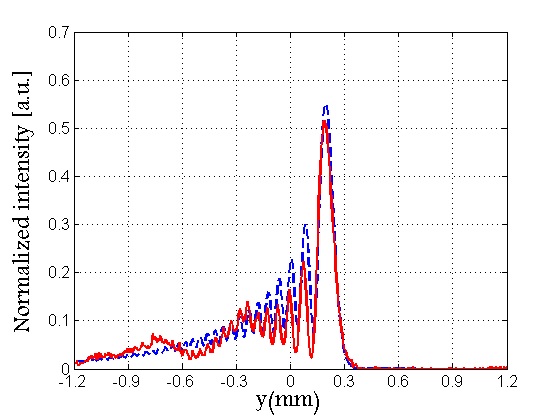}}
 \subfigure[]{\includegraphics[width=50mm]{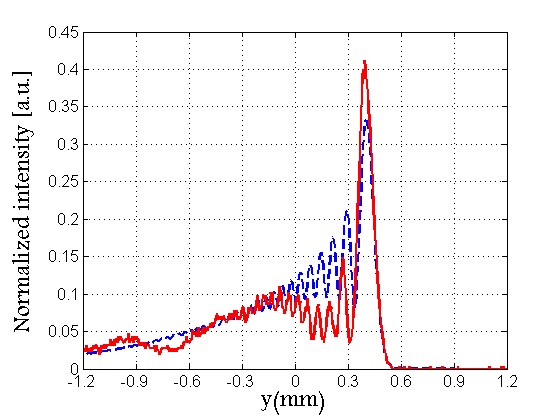}}
\caption{Transverse profile of normalized field for an Airy beam in $(1+1)$D theoretical (blue line) and experimental (line red) when $x_{0}=50 \mu m$ and $a=0.07$ in $(a)$ $z=3cm$, $(b)$ $z=7cm$, $(c)$ $z=11cm$, $(d)$ $z=15cm$.}
\label{Perfil_1D_50_0.07_1}
\end{figure}

\begin{figure}[H]
\centering
 \subfigure[]{\includegraphics[width=50mm]{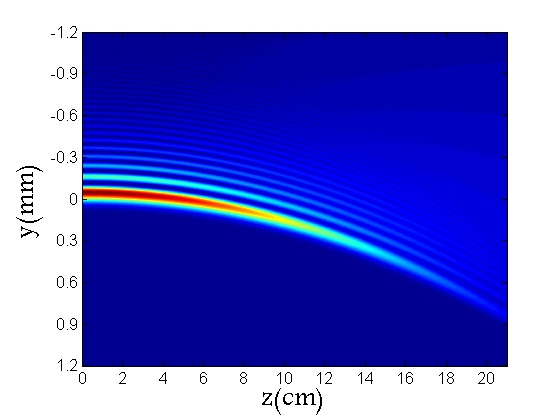}}
 \subfigure[]{\includegraphics[width=50mm]{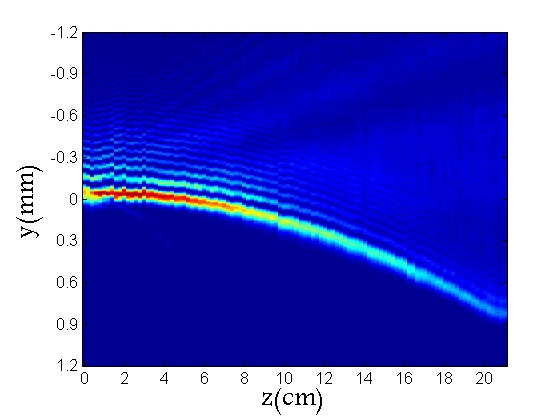}}
 \caption{Orthogonal projection of the dynamic propagation for an Airy beam in $(1+1)$D with $x_{0}=50 \mu m$ and $a=0.07$. (a) Theoretical description, (b) Experimental result.}
 \label{Dinamica_1D_50_0.14_2}
\end{figure}

Initially, we assume that the complex field that describes the Airy beam propagating with energy finite in $(1+1)$D and $(2+1)$D is given by the Eqs.\ref{Airy_finity1} and \ref{Airy_2D_finita} respectively. For the numerical generation of holograms we employ a software developed for CGH generation from the Eq. \ref{Transmissão} adopting the carrier of frequencies  $\eta=\xi=\Delta p/5$ for the plane wave of reference according to the bandwidth of SLM \cite{Arrizon2005}, where $\Delta p=1/\delta p$ is the bandwidth and $\delta p$ is the individual size of each pixel. Using the experimental setup shown in the Fig. \ref{arranjo} were reconstructed optically by several Airy beam types:

\textbf{First case}: For an Airy beam in $(1+1)$D with $x_{0}=50 \mu m$ and $a=0.07$, displacing the CCD camera on the propagation axis (z axis) and capturing the images, it obtained the transversal intensity pattern in a different plane along the direction of propagation Fig. \ref{Transversal_1D_50_0.07_1} and \ref{Perfil_1D_50_0.07_1}. The dynamic acceleration can be seen in the Fig.\ref{Dinamica_1D_50_0.14_2}, where the trajectory parabolic is evidenced as a result of the term $z^{2}/4k^{2}x_{0}^{4}$ of the equation \ref{Airy_finity1}, remaining almost diffraction free. 

\begin{figure}[H]
\centering
 \subfigure[]{\includegraphics[width=50mm]{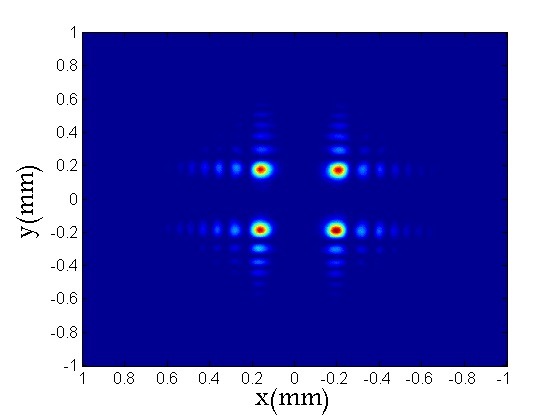}}
 \subfigure[]{\includegraphics[width=50mm]{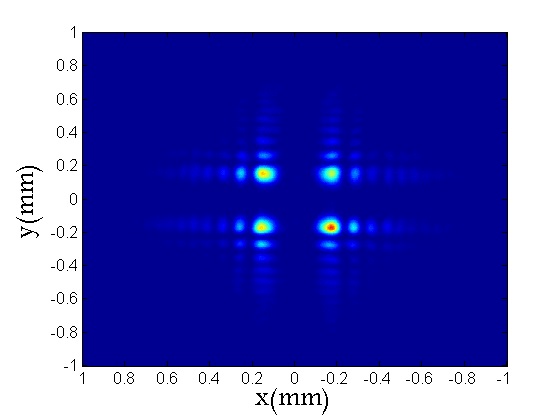}}
 \subfigure[]{\includegraphics[width=50mm]{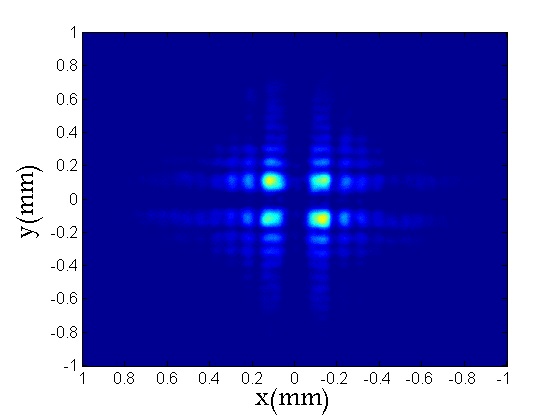}}
 \subfigure[]{\includegraphics[width=50mm]{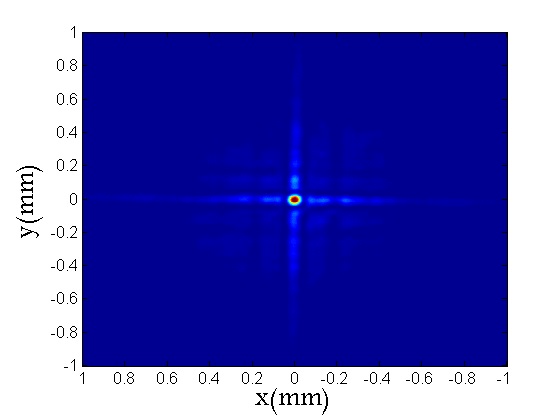}}
\caption{Evolution of the transverse intensity pattern for an array of 4 Airy beams in $(2+1)$D when $x_{0}=y_{0}=50 \mu m$, $\Delta x=\Delta y=0.15mm$ and $w_{1}=w_{2}=0.36mm$ that correspond to $a=0.14$ in  
$\left(a\right)$ $z=3cm$, $\left(b\right)$ $z=5cm$, $\left(c\right)$ $z=7cm$, $\left(d\right)$ $z=10cm$.}
 \label{Transversal_2D}
\end{figure}

\textbf{Second case}: In the same way, we reproduced experimentally an Airy beam in $(2+1)$D that propagates with finite energy. For the Eq.\ref{Airy_2D_finita} and choosing the same scales $x_{0}=y_{0}=50\mu m$ and $w_{1}=w_{2}$ that correspond to a truncation factor $a=0.07$ we can see in Fig.\ref{2D_0,07} the behavior of the transverse intensity pattern in different planes along the axis of propagation where the beam is accelerated in the same manner along the $45^{0}$ axis in the $x$-$y$ plane.
\begin{figure}[H]
\centering
 \subfigure[]{\includegraphics[width=50mm]{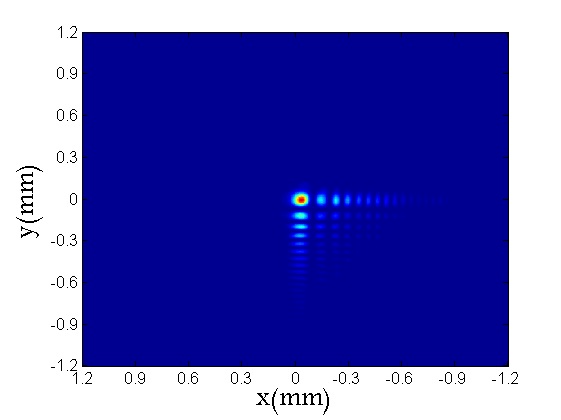}}
 \subfigure[]{\includegraphics[width=50mm]{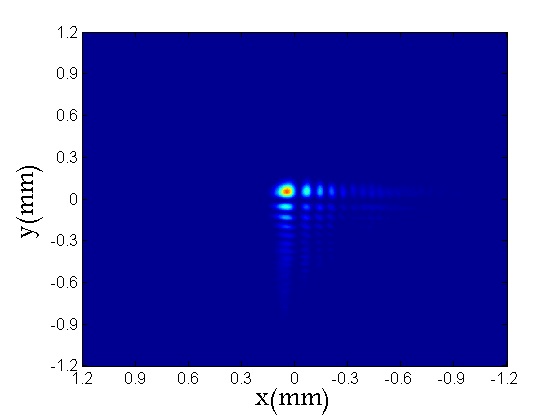}}
 \subfigure[]{\includegraphics[width=50mm]{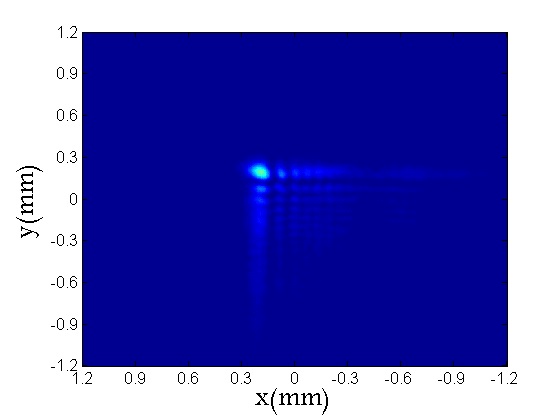}}
 \subfigure[]{\includegraphics[width=50mm]{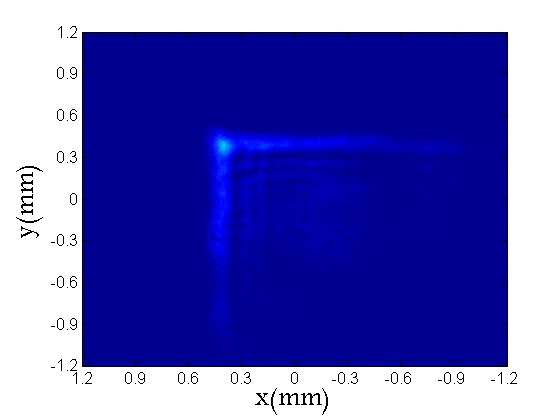}}
\caption{Transversal intensity pattern for an Airy beam in $(2+1)D$ when $x_{0}=y_{0}=50 \mu m$ and $w_{1}=w_{2}=0.71mm$ in $(a)$ $z=3cm$, $(b)$ $z=7cm$, $(c)$ $z=11cm$, $(d)$ $z=15cm$.}
 \label{2D_0,07}
\end{figure}
\textbf{Third case}: Process of numerical generation of Airy beam arrays in a single holographic reconstruction. Thus, considering the fields that define the beams $\phi\left(x,y,z \right)$, we take the origins of coordinates spatially displaced on the plane transversal in the direction of propagation $\left(\pm x+\bigtriangleup x, \pm y + \bigtriangleup y \right)$ we generate a CGH and from the transfer of the holographic pattern to the SLM it is reconstructed optically. For a set of $4$ Airy beams in $(2+1)D$ described by the fields \ref{Airy_2D_finita} and spatially displaced along the x and y axes, that is
\begin{equation}
\begin{aligned}
\phi_{1} & = \phi_{1}\left(x+\Delta x, y + \Delta y,z=0\right)\\
\phi_{2} & = \phi_{2}\left(-x+\Delta x, y + \Delta y,z=0\right)\\
\phi_{3} & = \phi_{3}\left(x+\Delta x, -y+\Delta y ,z=0\right)\\
\phi_{4} & =\phi_{4}\left(-x+\Delta x,-y+\Delta y,z=0\right)
\end{aligned}
\qquad
\label{Matriz_2_2D}
\end{equation}
using the field $\phi_{tot}=\sum_{i=1}^{4} \phi_{i}$ as an object beam to generate the CHG, we obtained the holographic reconstruction of the matrix of 4 identical Airy beams in $(2+1)$D with $x_{0}=y_{0}=50\mu m$, $\Delta x=\Delta y=0.15mm$ and $w_{1}=w_{2}=0.36mm$ that correspond to a truncation factor $a=0.14$, Fig.\ref{Transversal_2D}. 
The 4 beams are propagated along the $z$-axis and converge symmetrically at the same point due to the parabolic trajectory. In $z=10cm$ the beams undergo a deflection equal to the  initial displacement $(\Delta x=\Delta y=0.15mm)$. At this point, the intensity of the first central peak overlap forming in the origin of the $x$-$y$ plane a central maximum is well localized.
%
\section{Generation of Airy beams using photorefractive holography}
For the first time to our knowledge, photorefractive holography is being used to generate experimental Airy beams. The recording holographic in photorefractive crystal occurs via photorefractive effect which is a phenomenon where the refractive index of the medium is modulated by the incidence of a beam of light with spatial variation of the intensity. this modulation is given by $\Delta n =\left( -1/2\right)  n^{3} r_{13} E_{sc}$ in diffusive regime\cite{Gesualdi2006,Gesualdi2007,Gesualdi2008,Gesualdi2010,Brito2013}. The holographic reconstruction of a wave object (Airy beam) occur in quasi-real time, where the optical reconstruction is made through diffraction of the reference wave by a hologram recorded in the photorefractive crystal, diffracted wave. In BTO crystals we must consider the optical activity of the medium, this effect acts on the beam so as it propagates through the volume of the crystal polarization is rotated by $\theta_ao=\rho_{ao}L$ where $\rho_{ao}$ is the coefficient of optical activity. If $\lambda$ is the recording wave length, $L$ is the thickness, $m$ is the modulation of the incident interference pattern and $2\alpha$ is the angle between the interfering beams, the diffraction efficiency of photorefractive hologram is given by \cite{Gunter2006,Gesualdi2006,Gesualdi2007,Gesualdi2008,Gesualdi2010,Brito2013} :
\begin{gather}
\eta=\left(\frac{\pi\Delta n }{\lambda cos\alpha}\frac{sen\rho L}{\rho L} \right)^{2}m^{2} 
\label{eficiencia}
\end{gather}
\subsection{Experimental setup}
For holographic recording medium we used the BTO crystal $(Bi_{12}TiO_{20})$, of the silenite family, in diffusive regimen with dimensions ($9\times 8 \times 3 mm^3$ $L=3mm$ and electro-optical transverse configuration). This crystal has a lower band gap compared to other crystals, such as $(B_{12}SiO_{20})$ or $(B_{12}GeO_{20})$, which makes it suitable for wavelengths in the red region, as is the case of the laser He-Ne $(\lambda=632,8nm)$ used in our work. 

\begin{figure}[H]
 \centering
  \includegraphics[scale=0.37]{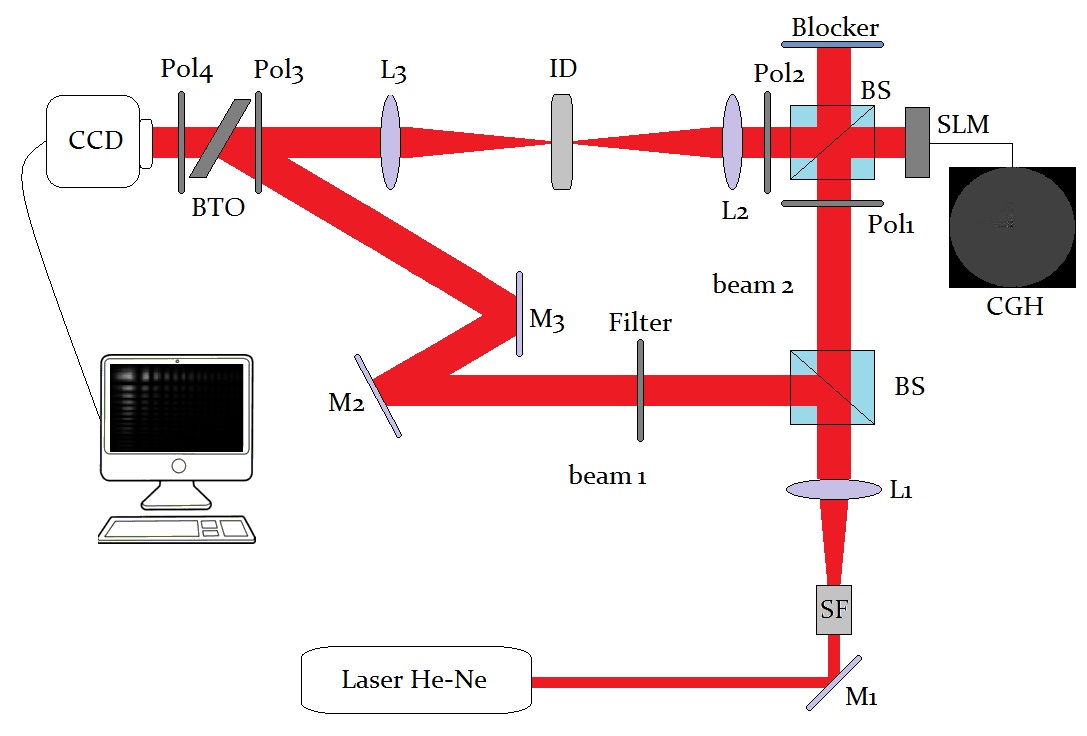}
\caption{Photorefractive holographic setup for optical generation of Airy beam, where Ms are mirrors, BS is a beam-splitter, FS is the spatial filter, L0s are lenses, Pol’s are polarizers, SLM is spatial light modulator, ID is the mask, BTO is the photorefractive crystal, Filter is neutral density filter and CCD camera for image acquisition. }
 \label{arranjo_fotorrefractivo}
\end{figure}

For the record and optical reading of Airy beams holograms, we set up an experimental arrangement of photorefractive holography shown in Fig. \ref{arranjo_fotorrefractivo}. The light reflected (beam 1) is used as a reference beam and the light transmitted (beam 2) is used as a object beam, the object beam (Airy beam) is generated optically from the SLM. The two beams propagating in the direction of the crystal arrive forming an angle of approximately $30^{0}$. In addition, due to the optical activity of the BTO crystal $(\rho_{ao}=6.9^{o}mm^{-1})$, the reference beam and beam object should incise on the surface of the crystal with a polarization angle of $10^{0}$ with the vertical for at its center  both beams are vertically polarized and the interference pattern is maximized. This choice of polarization direction is made through the polarizing placed in the crystal input $P_{3}$. 
The combination of the object beam with the reference beam generates a pattern of interference that modulates the refractive index of the crystal via electro-optical effect, allowing the recording of the hologram in the crystal. For the reconstruction of the hologram, the object beam is blocked and the crystal output is positioned a polarized $P_{4}$ that blocks the transmitted beam allowing the passage of the diffracted beam that contains information about the hologram, the holographic image. Finally, the reconstructed beam is recorded by the CCD camera and transmitted for visualization, storage and processing in the computer.
\subsection{Experimental results}
\begin{figure}[H]
\centering
 \subfigure[]{\includegraphics[width=50mm]{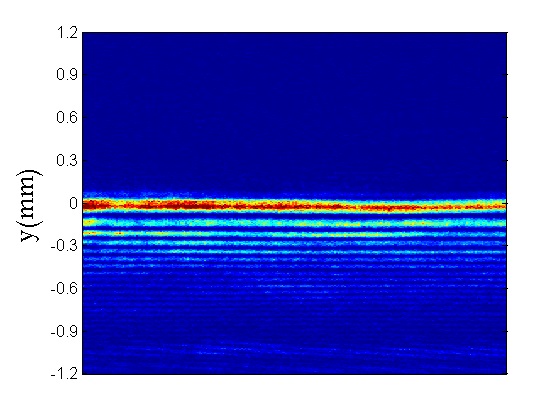}}
 \subfigure[]{\includegraphics[width=50mm]{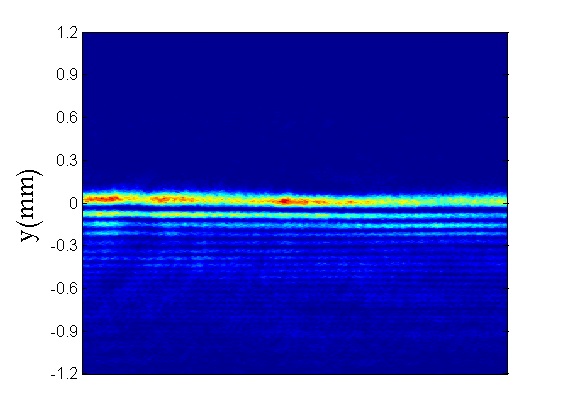}}
 \subfigure[]{\includegraphics[width=50mm]{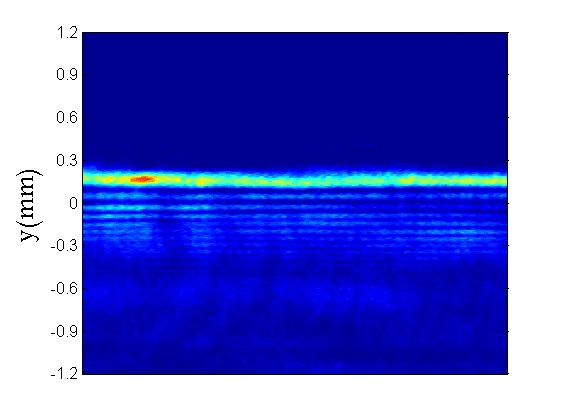}}
 \subfigure[]{\includegraphics[width=50mm]{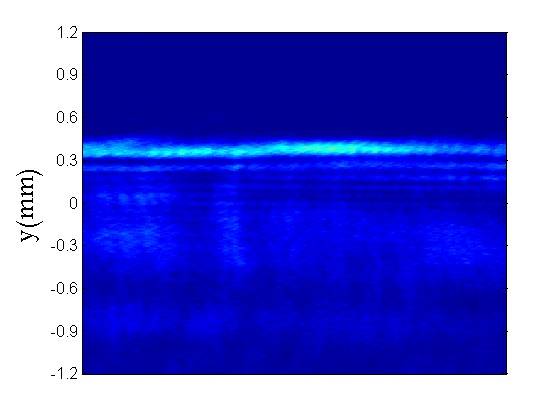}}
\caption{Transversal intensity pattern for an Airy beam in $(1+1)D$ reconstructed using photorefractive holography when $x_{0}=50 \mu m$ and $a=0.07$ in $(a)$ $z=3cm$, $(b)$ $z=7cm$, $(c)$ $z=11cm$, $(d)$ $z=15cm$.}
 \label{Airy_1D_BTO}
\end{figure}

\begin{figure}[H]
\centering
 \subfigure[]{\includegraphics[width=50mm]{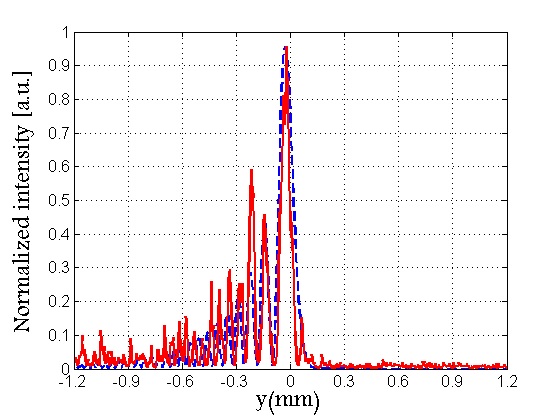}}
 \subfigure[]{\includegraphics[width=50mm]{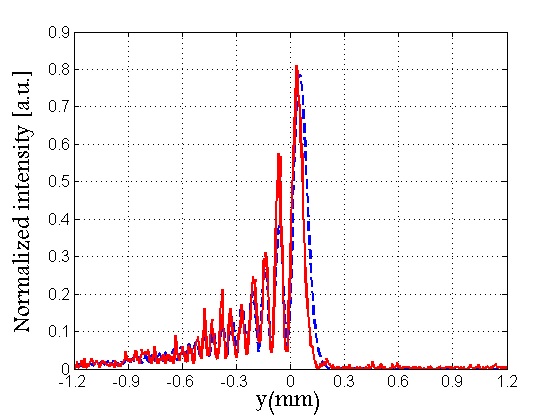}}
 \subfigure[]{\includegraphics[width=50mm]{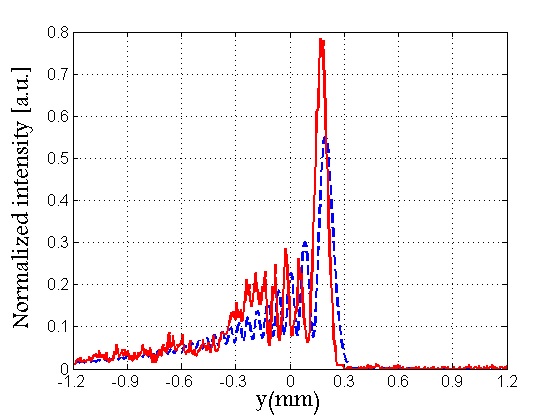}}
 \subfigure[]{\includegraphics[width=50mm]{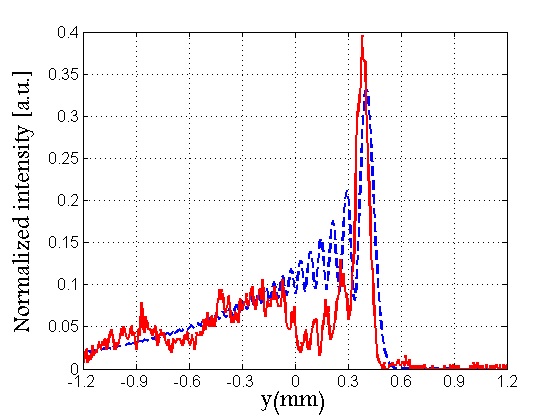}}
\caption{Transverse profile of normalized field for an Airy beam in $(1+1)$D theoretical (blue line) and experimental reconstructed using photorefractive holography (line red) when $x_{0}=50 \mu m$ and $a=0.07$ in $(a)$ $z=3cm$, $(b)$ $z=7cm$, $(c)$ $z=11cm$, $(d)$ $z=15cm$.}
 \label{Dinamica_1D_50_0.07_1}
\end{figure}

The interference pattern formed by the Airy beam (object beam) and the reference beam are incident on the 
BTO crystal surface until the space charge field is maximum and the holographic grating reaches the saturation level (approximately 30 seconds). Then, the object beam is blocked and only the reference beam is focused on the crystal, the reading process, which is diffracted reconstructing the photorefractive hologram and the diffracted beam reconstructed as an Airy beam.
\textbf{First case}: Airy beam in $(1+1)$D with $x_{0}=50 \mu m$ and $a=0.07$.  The Airy beam is generated by the SLM and recorded optically in the photorefractive crystal. The Figs. \ref{Airy_1D_BTO} and \ref{Dinamica_1D_50_0.07_1} shows the holographic reconstruction of the transversal intensity pattern in a different plane along the direction of propagation.
%

\begin{figure}[H]
\centering
 \subfigure[]{\includegraphics[width=50mm]{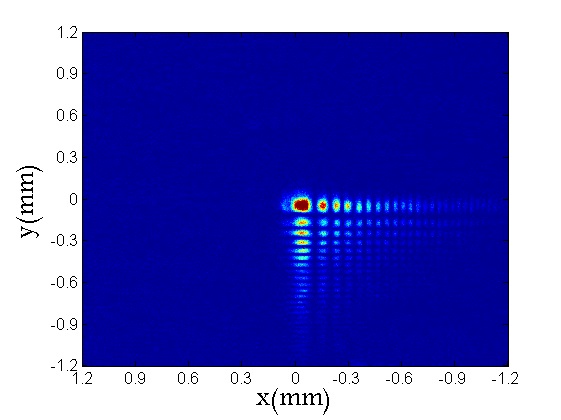}}
 \subfigure[]{\includegraphics[width=50mm]{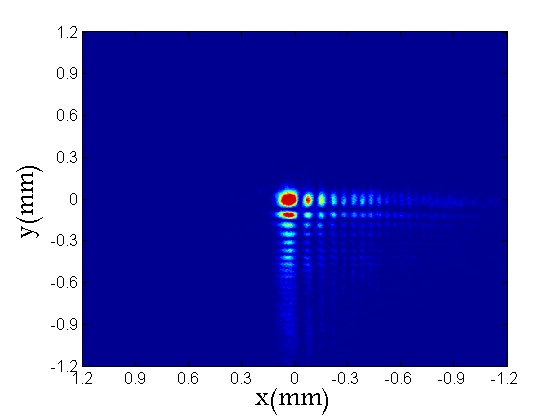}}
 \subfigure[]{\includegraphics[width=50mm]{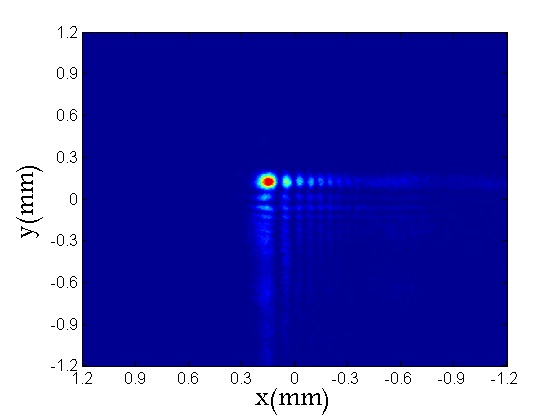}}
 \subfigure[]{\includegraphics[width=50mm]{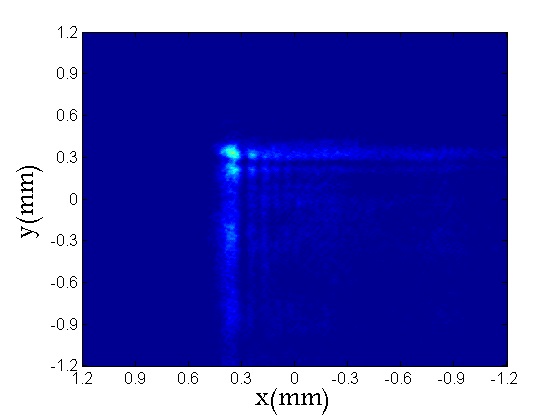}}
\caption{Transversal intensity pattern for an Airy beam in $(2+1)D$ reconstructed using photorefractive holography when $x_{0}=y_{0}=50 \mu m$ and $w_{1}=w_{2}=0.71mm$ in $(a)$ $z=3cm$, $(b)$ $z=7cm$, $(c)$ $z=11cm$, $(d)$ $z=15cm$.}
 \label{Airy_2D_BTO}
\end{figure}
\begin{figure}[H]
\centering
 \subfigure[]{\includegraphics[width=55mm]{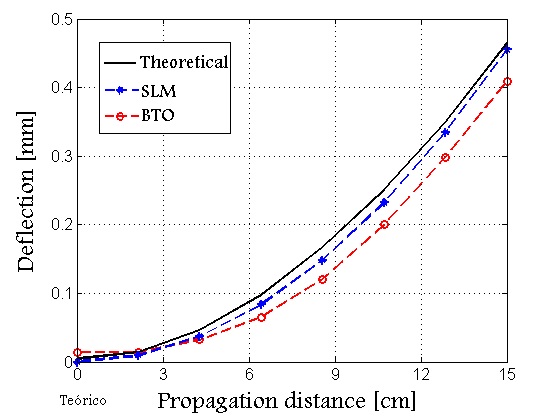}}
 \subfigure[]{\includegraphics[width=55mm]{Deflexao_50_0,07_BTO_1.jpg}}
\caption{Comparison of the transversal acceleration as a function of propagation distance for an Airy beam:  predicted theoretically, generated by SLM and reconstructed via holographic photorefractive in $(a)$ $(1+1)$D with $x_{0}=50 \mu m$ and $a=0.07$, $(b)$ $(2+1)$D with $x_{0}=y_{0}=50 \mu m$ and $w_{1}=w_{2}=0.71mm$ corresponding to $a=0.07$.}
 \label{Deflexao}
\end{figure}
\begin{figure}[H]
\centering
 \subfigure[]{\includegraphics[width=50mm]{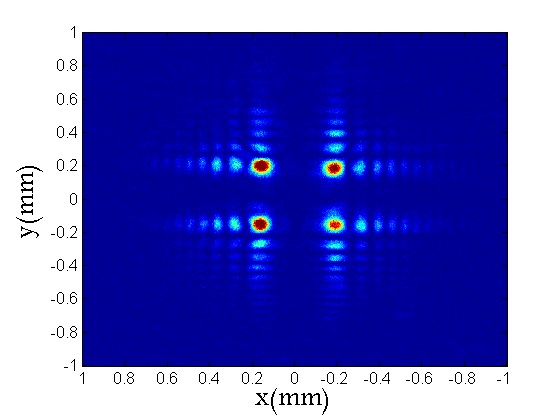}}
 \subfigure[]{\includegraphics[width=50mm]{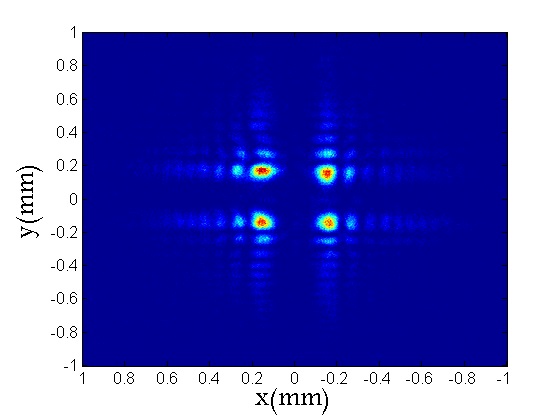}}
 \subfigure[]{\includegraphics[width=50mm]{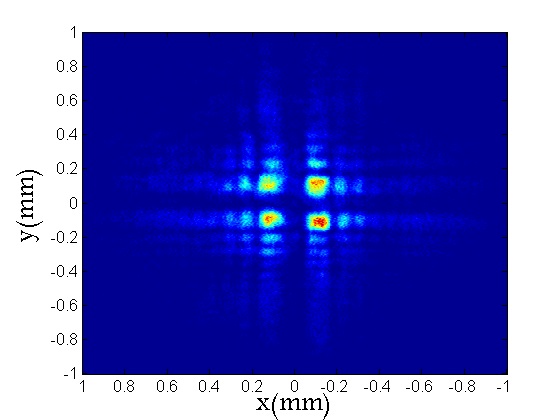}}
 \subfigure[]{\includegraphics[width=50mm]{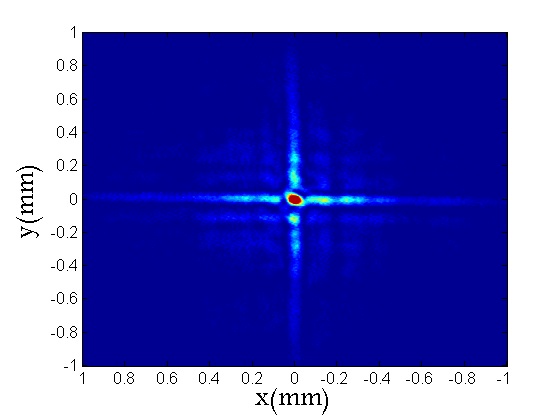}}
\caption{Evolution of the transverse intensity pattern for an array of 4 Airy beams reconstructed using photorefractive holography in $(2+1)$D when $x_{0}=y_{0}=50 \mu m$, $\Delta x=\Delta y=0.15mm$ and $w_{1}=w_{2}=0.36mm$ that correspond to $a=0.14$ in  
$\left(a\right)$ $z=3cm$, $\left(b\right)$ $z=5cm$, $\left(c\right)$ $z=7cm$, $\left(d\right)$ $z=10cm$.}
 \label{Matriz_BTO_2D}
\end{figure}

\textbf{Second case}: Similarly, we reproduce experimentally an Airy beam in $(2+1)$D. The hologram is characterized by the same scales x-y and the parameters $x_{0}=y_{0}=50 \mu m$ and $w_{1}=w_{2}=0.71mm$ corresponding to $a=0.07$. The results of the optical reconstruction are shown in Fig \ref{Airy_2D_BTO}, where we see the evolution of the transverse intensity pattern on the propagation axis.  
The Figures \ref{Deflexao}(a) and (b) shows the transversal acceleration as a function of the distance of propagation for an Airy beam in $(1+1)$D and $(2+1)$D respectively, where the parabolic trajectory is described by the theoretical relation $\lambda^{2}z^{2}/16\pi^{3}x_{0}^{3}$. The black line corresponds to the theoretical prediction, blue line to optical reconstruction using SLM and red line to optical reconstruction in the photorefractive medium. 
\textbf{Third case}: Process of optical reconstruction of an Airy beam array in a single holographic reconstruction. Using the experimental arrangement shown in the fig.\ref{arranjo_fotorrefractivo}, and the same way as in the case of the single beam in $(1+1)$D and $(2+1)$D, We reconstructed optically the holograms arrays recorded in crystal. In the Fig \ref{Matriz_BTO_2D} shows the holographic reconstruction of the transverse profile of intensity in different planes along the direction of propagation for an array of 4 Airy beams when $x_{0}=y_{0}=50 \mu m$, $w_{1}=w_{2}=0.43 $ corresponding to $a=0.14$ and spatially displaced by $\Delta x =\Delta y= 0.15mm$. 
We see that the Airy beams reconstructed by a photorefractive holographic system are in great agreement with those obtained with the spatial light modulator SLM and consequently with theoretical expected outcome. In the BTO crystal, the process of recording and reconstruction occurs dynamically presenting a fast response time for the formation of the holographic network and a high rate of recombination for the quenching of it. Thus, it is interesting to note that for the diffusion regimen and despite the low diffraction efficiency present in the BTO crystal it was possible to reconstruct from the block of object beams the diffracted beam with a good contrast on the propagation axis. One advantage of the generation of Airy beam using the photorefractive crystal in relation to the spatial light modulators is the direct generation of the Airy beam without the need of a 4-f system, furthermore, the photorefractive crystals have a high resolution which is greater than any liquid crystal display.
\section{Conclusions}
We validate the experimental procedure of the computational and photorefractive holography for optical generation of Airy beams. 

The implementation of amplitude CGHs over LC-SLM for generation of Airy beam in $(1+1)$D and $(2+1)$D with finite energy, where the holographic recording is numerical and the reconstruction holographic is optical, presented good results. In addition, due to the process of construction numerical of the fields and holograms it was possible to create arrays of Airy beams single reconstruction holographic. 

The photorefractive holography, the most interesting, presented results where the beams have the same non-diffractive properties initially obtained with the SLM and with the theory. This means that photorefractive holographic  preserves the information of the Airy beams and is present as a new  possibility of expermental generating of Airy beams and/or Airy beams arrays. 

The experimental results are in agreement with the theoretical predictions, and open exciting possibilities of generating many further potentially interesting Airy beams and/or Airy beam arrays for scientific and technological applications: optical micromanipulation, plasma physics, optical microscopy, optical communications and optical vortices on Airy beams. 

\section*{Acknowledgment}
We thank Michel Zamboni-Rached for many stimulating discussions. This research is supported by the UFABC, CAPES, FAPESP (grant 09/11429-2) and CNPQ (grants 476805/2012-0 and 313153/2014-0).
 
\section*{References}

\end{document}